# "Cosine directions using Rao-Blackwell Theorem and Hausdorff metric in Quasars"

Byron E. Bell
DePaul University
Chicago, IL
bbell2851@gmail.com

**Abstracts**
This analysis will determine the equations of the cosine directions for all flux of the optical spectrum in quasars. Studies on Hausdorff metric will greatly enhance our understanding of quasars distances. This study will complete steps in the classification of quasars by finding the minimum variance of flux by using the Rao–Blackwell Theorem. The papers of C. R. Rao and D. Blackwell will be examined to clarify more of the above theorem.

## 1. Introduction

The study of cosine directions using the Rao-Blackwell theorem and Hausdorff metric is needed for the study of quasars. The cosine directions of quasars have been used before by this author (Bell 2010). In this study will occur by using apparent magnitude data transforming it into the rate of energy transfer per unit area and he after will be called just flux. Flux is measured as Watts per square meter (W·m$^{-2}$) per second. The theory of flux will be looked at by cosine directions from a deterministic standpoint and then apply mathematical statistics to the cosine directions equations. Hausdorff metric, Henrikson (1999) will be utilized to find distances between quasars from applying mathematical analysis to flux. The Rao–Blackwell Theorem is from papers by C. R. Rao (1945, 1946) and David Blackwell (1947). The first paper of the trio shows the derivation of expectation and variance, and then he explains, from the mathematics of statistics standpoint the Einstein theory of 1916 using distance in three forms and finding distance by the use of population perimeters. The work of Rao and Blackwell in distance studies are quite amazing, given now modern computers and software, now their work can be more fully utilized and viewed. The Hausdorff metric as well is great tool to have in this new field of astrostatistics in analyzing distance data as well. Possibly this could lead to a more formal picture of quasars. Data for this study comes from SDSS Quasar Catalog V, Seventh Data Release (Schneider et al, 2010). Data that is used was created in a MS Excel data set.

## 2. Results: Theory of Flux

Flux values are made from having $m_o=0$ and $f_o=1$, and getting the first estimation of the column(s) of data from this process and the flux equation of $f=2*10^{-2m_o/5} *\{f_o \sinh[(2\ln(10)/5)(m_o-m_{ugriz})]\}$ from the estimation of the column(s) of data a second column(s) flux data is created this is what will be used to make models from and come up with conclusions of flux. Models will be made are linear and nonlinear regression equations. Then general model of using the second approximation of the flux data to find a mean of flux would become $f_{ugriz, hat}= \alpha + \alpha_u f_u + \alpha_g f_g + \alpha_r f_r + \alpha_i f_i + \alpha_z f_z + \varepsilon$. For the general form of the flux data would model the statistics and without a constant term and no error term, $f_{ugriz} = \alpha_u f_u + \alpha_g f_g + \alpha_r f_r + \alpha_i f_i + \alpha_z f_z$. Where a model has no constant term and an error term equation, $f_{ugriz,hat}=\alpha_u f_u + \alpha_g f_g + \alpha_r f_r + \alpha_i f_i + \alpha_z f_z + \varepsilon$ would mimic the actual data in different manners from the above models. Following the work of (Bell, 2010) the cosine directions will be used to give new models and equations that predict flux in all spectrums of energy or flux optical levels. Then the overall forms of new calculations will be the following equations or models a third approximation of the flux data, $F_{ugriz,hat} = \alpha_u F_u + \alpha_g F_g + \alpha_r F_r + \alpha_i F_i + \alpha_z F_z + \varepsilon$ is using cosine directions. Then view the data with a error and constant terms $F_{ugriz,hat} = \alpha + \alpha_u F_u + \alpha_g F_g + \alpha_r F_r + \alpha_i F_i + \alpha_z F_z + \varepsilon$. The model will be examine without error and constant terms in the follow general equation $F_{ugriz,hat} = \alpha_u F_u + \alpha_g F_g + \alpha_r F_r + \alpha_i F_i + \alpha_z F_z$, in cosine directions linear and nonlinear regression methods. Examining the some of the above data gives an regression results of the following: the second approximation of the flux data and coefficients of flux $f_u, f_g, f_r, f_i, f_z$ of the following set of data variables;(standard error) are $f_{ugriz}$;(SE): 0.416446382;(0.001987806), -0.14673133;(0.006080998),0.181200438;(0.010446884) ,-0.728604144;(0.01105773), 0.299987556;(0.007590644) the sample size (n) of 105,783 data points over five (5) columns for forecasting redshift (z). The Regression Statistics are the following R-Square is 0.879057612 and the Adjusted R-Square is 0.879043585 of the statistical model in addition to the Multiple R of 0.937580723 and the standard error is 0.616028474. From the ANOVA the F is 153765.903, P-value is .0001. In this analysis, the F-value corresponds to a P-value or (Significant) Sig of .0001. The hypothesis F-test, $H_0: \rho_{ugriz}=0$, $H_1:\rho_{ugriz}\neq 0$, $\alpha > Sig. \rightarrow$ Reject $H_0$, $\alpha < Sig. \rightarrow$ Accept $H_0$, .05>0$\rightarrow$ Reject $H_0$. The decision is to reject $H_0$ at $\alpha=.05$ level. Therefore it is a statistically significant relationship that exists between the dependent variable redshift (z) and independent variables of flux $f_u, f_g, f_r, f_i, f_z$. Then compare this to the regression results of cosine directions of flux $F_u, F_g, F_r, F_i, F_z$ where redshift (z) is the outcome variable. Following set of data, variables; (standard error) are $F_{ugriz}$;(SE):-29.81212443; (0.142817556), 10.66147473;(0.44481049),-12.21898782;(0.765559688), 53.67447528; (0.807606683),-21.12873951;(0.553547485) the sample size (n) of 105,783 data points over five (5) columns for forecasting redshift (z). The Regression Statistics are the following R-Square is 0.875302145 and the Adjusted R-Square is 0.875287975 of the statistical model in addition to the Multiple R of 0.935575836 and the standard error is 0.625496262. From the ANOVA the F is 148497.8785, P-value is .0001. In this analysis, the F-value corresponds to a P-value or (Significant) Sig of 0. The hypothesis F-test, $H_0: \rho_{ugriz}=0$, $H_1:\rho_{ugriz}\neq 0$, $\alpha > Sig. \rightarrow$ Reject $H_0$, $\alpha < Sig. \rightarrow$ Accept $H_0$, .05> .0001$\rightarrow$Reject $H_0$. The decision is to reject $H_0$ at $\alpha=.05$ level. Therefore it is a statistically significant relationship that exist between the dependent variable redshift (z) and independent variables of Cosine directions flux, $F_u, F_g, F_r, F_i, F_z$. Likewise the statistical analysis of Right Ascension (R.A) and actual flux $f_u, f_g, f_r, f_i, f_z$ and then the Right Ascension (R.A.) and the Cosine directions flux $F_u, F_g, F_r, F_i, F_z$. it is also a statistically significant relationship as well. The work of (Lupton et al, 1999) and (Mortlock et al, 2012) lay the foundation on" Theory of Flux "of this paper. What is clearly has been seen that redshift (z) and Right Ascension (R.A.) are outcome variables and a causal factor(s) of $F_{ugriz}$.

### 3. Hausdorff metric

The Hausdorff metric will be used to show distances between quasars utilizing flux energy. A proof will be use to show the basics of the Hausdorff metric or distance from a mathematical deterministic standpoint. Then the Hausdorff metric will be use to further enhance our understanding of distances between quasars. $d_{HM}(x_u, x_g) = Max \{Sup_{f_u \varepsilon x_u} Inf_{f_g \varepsilon x_g} d(f_u, f_g), Sup_{f_g \varepsilon x_g} Inf_{f_u \varepsilon x_u} d(f_u, f_g)\}$, Henrikson (1999), $x_u=\{f_{1u}, f_{2u},...,f_{nu}\}$, $x_g=\{f_{1g}, f_{2g},...,f_{ng}\}$, $x_r=\{f_{1r}, f_{2r},...,f_{nr}\}$, $x_i=\{f_{1i}, f_{2i},...,f_{ni}\}$, $x_z=\{f_{1z}, f_{2z},...,f_{nz}\}$, $d=|f_{griz} - f_u|$ of $R^{ugriz}$, generalize equation of distance, $d(f_u, f_z)=((f_g-f_u)^2 + (f_r-f_u)^2 + (f_i-f_u)^2 + (f_z-f_u)^2)^{1/2}$. Data points that is next to one another or columns that contain data points that is next to one another. The following formula, $d=|f_{griz} - f_u|$ is individualized data point distances of flux. Given the following let the distance of the quasars from ultraviolet light (u) to even more red light (z), is given as a proof : then $|f_u-f_g| + |f_g-f_r| + |f_r-f_i| + |f_i-f_z| = |f_u-f_g + f_g-f_r + f_r-f_i + f_i-f_z|$

∴ | $f_u$-$f_z$|,( Huttenlocher, D. P., et al. 1993). Hence will we want to know the distance between quasars that are next to one another in space we can look at the individual distances of quasars as the following: |$f_u$-$f_g$|, |$f_g$-$f_r$|, |$f_r$-$f_i$|, |$f_i$-$f_z$|, this comes out as the optical spectrum of light otherwise known as apparent magnitude and energy which we measure as flux $f_{ugriz}$. The Hausdorff metric will be shown as the following: $d_{HM}(x_u, x_g)$= Max{Sup $f_u$ ε $x_u$ Inf $f_g$ ε $x_g$ d($f_u$, $f_g$), Sup $f_g$ ε $x_g$ Inf $f_u$ ε $x_u$ d($f_u$, $f_g$)}=Max{d(60.806, 44.622),d(44.760, 60.386)=Max{16.184,15.626 }=16.184 as a Hausdorff metric or distance between the ultraviolet flux and the green flux is 16.184 watts per second meter squared(W/m$^2$). To continue with the analysis that examined the other spectrums of flux from the green spectrum (g) to the even more red light spectrum (z) of flux in those quasars of the 105,783 population of this study. $d_{HM}(x_g, x_r)$=15.884W/m$^2$, $d_{HM}(x_r, x_i)$=12.15W/m$^2$, $d_{HM}(x_i, x_z)$= 11.857W/m$^2$. The distance analysis of flux energy from quasars can be summed up as the following: The entire energy spectrum of quasars are decreasing from ultraviolet light (u) to even more red light (z). The spectrum from ultraviolet light (u) to red light (i) is decreasing at a rate of .3 W/m$^2$, the spectrum from green light (g) to more red light (i) is decreasing at a rate of 3.734 W/m$^2$, the spectrum from red light(r) to even more red light(z) is decreasing at a rate of .293W/m$^2$.Regression analysis of Cosine directions of flux data and coefficients of flux $F_u$, $F_g$ ,$F_r$, $F_i$, $F_z$ of the following set of data variables;(standard error) $F_{ugriz}$ ; (SE): 0.718512841; (0.028133845), -3.196474611; (0.087632587), 3.665354745;(0.150825183), -44.13642513; (0.15910341), 43.0810976;(0.109059071) sample size (n) of 105,783 data points over five (5) columns for forecasting |$f_i$ –$f_z$|. The Regression Statistics are the following R-Square is 0.814881335 and the Adjusted R-Square is 0.814864881 of the statistical model in addition to the Multiple R of 0.902707779 and the standard error is 0.123227502. From the ANOVA the F is 93124.81031, P-value is .0001. In this analysis, the F-value corresponds to a P-value or (Significant) Sig of .0001. The hypothesis F-test, $H_{0:}$ $ρ_{ugriz}$=0, $H_1$:$ρ_{ugriz}$≠0, α >Sig.→ Reject $H_0$, α < Sig.→ Accept $H_0$, .05>0→ Reject $H_0$.The decision is to reject $H_0$ at α=.05 level. A statistically significant relationship that exist between the dependent variable |$f_i$ –$f_z$| and independent variables of Cosine directions flux $F_u$, $F_g$ ,$F_r$, $F_i$, $F_z$. Cosine directions of flux $F_u$ and coefficient of $F_u$, of the following set of data variables; standard error (SE) is 54.35174767, coefficients of flux $F_u$ is -57.07295116; SE is (0.062067765), sample size (n) of 105,783 data points over one (1) columns for forecasting |$f_u$ –$f_g$|. R-Square is 0.888805831 and the Adjusted R-Square is 0.88880478 of the statistical model in addition to the Multiple R of 0.942764993 and the SE is 0.363092898. From ANOVA the F is 845528.9, P-value is 0. In this analysis, hypothesis F-test, $H_0$:$ρ_{u-g}$=0,$H_1$:$ρ_{u-g}$≠0,.05>0→ Reject $H_0$.The decision is to reject $H_0$ at α=.05 level. A statistically significant relationship exist between the dependent variable |$f_u$ –$f_g$| and independent variable of Cosine direction flux $F_u$. Hausdorff metric is a powerful tool that could be used in astrophysics or astrostatistics in measuring distances in space.

**4. Rao–Blackwell Theorem**

The study of quasars need to be further advanced by the use of techniques, processes, methods and models of the Rao–Blackwell Theorem. In this section of the paper I will show the formula and demonstrate statistical analysis from distances equation(s) perspective. Rao (1945) work on distance formulas and equations is groundbreaking using this format from Dr. Einstein we can obtain processes of attaining distances in space and nonlinear determination of statistical distances. Rao (1945) work on distance is in three forms the first form is where the distances between two objects where the slopes are not equal ($m_u$ ≠ $m_{griz}$) and the standard deviations are not statistically significant ($σ_u$ ≠ $σ_{griz}$). This is calculated by unique equation of distance that Rao developed using trigonometric formula. The aforementioned equation is using angles from the two populations that we have data from. The second form to calculate distances developed by

Rao (1945), is where the slope of the two populations are the same or statistically significant ($m_u = m_{griz}$) and the standard deviation of the two populations are not statistically significant ($\sigma_u \neq \sigma_{griz}$). This equation is utilizing logarithms to measure the distances between quasars measured in watts per meter squared. The last and final form to calculate distance based on Rao (1945) work is where the slope of the first population and the slope of the second population are not equal or not statistically significant ($m_u \neq m_{griz}$), and the standard deviations of the two populations are equal or has statistically significant to one another ($\sigma_u = \sigma_{griz}$). In the first form $\theta_{ugriz}=\cos^{-1}(f_{ugriz}/a)$ where a is the radius of the flux the author use this form to make the calculation easier than the original equation of Rao (1945). From this we can establish distances based on all three forms from the data we have obtain. By using regression analysis we can establish statistically significant or not statistically significant relationships between distances. This would be most helpful in knowing levels of watts per meter squared measurements of quasars.

The first form of the distance equation:
If $m_u \neq m_{griz}$ and $\sigma_u \neq \sigma_{griz}$ then the distance comes out as

$$D_{ug}=\sqrt{2}\ln[\tan(\theta_u/2)/\tan\theta_g/2)]$$

where

$$\theta_{ugriz}=\cos^{-1}(f_{ugriz}/a) \text{ where a is the radius of the flux}$$

The second form of the distance equation:
If $m_u = m_{griz}$ and $\sigma_u \neq \sigma_{griz}$

$$D_{ug}=\sqrt{2}\ln[(\sigma_u/\sigma_g)]$$

The third form of the distance equation:
If $m_u \neq m_{griz}$ and $\sigma_u = \sigma_{griz}$

$$D_{ug}=|m_u-m_g|/\sigma$$

where $\sigma=(\sigma_u+\sigma_g)/2$

This work will give the results of the distance equation(s) to first form:

| $D_{ug}=\alpha_u*F_u$ | $D_{gr}=\alpha_g*F_g$ | $D_{ri}=\alpha_r*F_r$ | $D_{iz}=\alpha_i*F_i$ |
|---|---|---|---|
| (se$_u$) | (se$_g$) | (se$_r$) | (se$_u$) |
| $D_{ug}=1.95037*F_u$ | $D_{gr}=1.99492*F_g$ | $D_{ri}=2.01394*F_r$ | $D_{iz}=2.02668*F_i$ |
| (2.18E-04) | (7.14E-05) | (9.98E-05) | (1.14E-04) |

For all four equations the $R^2=.999$ and Adj $R^2=.999$, at the $\alpha=.05$ level, P-Value is .0001
For all four equations the coefficient of the variables are statistically significant.
The variances of the distances are shown as the following: $S^2(D_{ug})=1.38\text{E-}0303W/m^2$ ,$S^2(D_{gr})=2.91\text{E-}0403W/m^2$ , $S^2(D_{ri})=1.02\text{E-}0403W/m^2$ , $S^2(D_{iz})=9.84\text{E-}0503W/m^2$. Based on the Rao-Blackwell Theorem the distance the between optical light of i and the optical light of z, is the distance that has the lease errors in that spectrum , $S^2(D_{ug})$ of $1.38\text{E-}03W/m^2$.

## 3. Conclusions

The use of flux as a tool to measure distance is much needed in astrostatistics. The Hausdorff metric as a mathematical measure of distance is another way to measure quasars and other astronomical phenomenon in space. Rao–Blackwell Theorem on distances is indispensable to the study of astrophysics/astrostatistics with this knowledge of using the work of Rao (1945) on distances and of the lease variances can be use in the future for classification of quasars and other data of astronomy.

**References**

Bell, B. E. (2010). The 1905 Einstein equation in a general mathematical analysis model of Quasars. *ADA 6 - Sixth Conference on Astronomical Data Analysis*.

Blackwell, D. (1947). Conditional expectation and unbiased sequential estimation. *Annals of Mathematical Statistics, 18*(1), 105-110.

Henrikson, J. (1999). Completeness and Total Boundedness of the Hausdorff Metric. *MIT Undergraduate Journal of Mathematics*(1), 69-80.

Huttenlocher, D. P. , et al. (1993). Comparing Images Using The Hausdorff Distance. *IEEE*, 850-854.

Lupton, R. , et al. (1999). A Modified Magnitude System that Produces Well-Behaved Magnitudes, Colors, and Errors Even for Low Signal-to-Noise Ratio Measurements. *The Astronomical Journal, 118*(3), 1406-1410.

Mortlock, D. J. , et al. (2012). Probabilistic selection of high-redshift quasars. *Monthly Notices of the Royal Astronomical Society, 419*, 390–410.

Rao, C. (1945). Information and accuracy attainable in the estimation of statistical parameters. *Bulletin of Calcutta Mathematical Society, 37*(3), 81-91.

Rao, C. (1946). Minimum variance and the estimation of several parameters. *Cambridge Philosophical Society, 43*(2), 280-286.

Schneider, D. P., et al. (2010). The Sloan Digital Sky Survey Quasar Catalog. V. Seventh Data Release. *The Astronomical Journal, Volume 139, Issue 6,*1-37.